\documentclass[conference]{IEEEtran}

\usepackage[utf8]{inputenc}
\usepackage{cite}
\usepackage[draft]{hyperref}
\usepackage{amsmath,amssymb,amsfonts}
\usepackage{amsthm}
\usepackage{enumitem} 
\usepackage{braket}

\usepackage{booktabs}

\usepackage{tikz}
\usetikzlibrary{arrows,shapes,snakes}
\usetikzlibrary{calc}
\usetikzlibrary{automata}
\usetikzlibrary{patterns}

\definecolor{fg0}{RGB}{33,113,181}
\definecolor{fg0medium}{RGB}{116,169,207}
\definecolor{fg0light}{RGB}{189,215,231}
\definecolor{fg1}{RGB}{49,163,84}
\definecolor{fg2}{RGB}{49,163,84}
\definecolor{hghlght}{RGB}{127,205,187}

\usepackage[textsize=scriptsize]{todonotes}
\renewcommand{\todo}[1]{} 

\DeclareMathOperator{\tCore}{core}
\DeclareMathOperator{\tPrio}{prio}
\DeclareMathOperator{\tBceT}{bcet}
\DeclareMathOperator{\tDead}{deadline}
\DeclareMathOperator{\tPree}{preemptable}
\DeclareMathOperator{\tActT}{act\_type}

\DeclareMathOperator{\tOffset}{offset}
\DeclareMathOperator{\tPeriod}{period}
\DeclareMathOperator{\tPredecessor}{predecessor}
\DeclareMathOperator{\tMaxDeltaT}{\Delta t\_max}
\DeclareMathOperator{\tMinDeltaT}{\Delta t\_min}
\DeclareUnicodeCharacter{00A0}{ hier}

\title{Estimating Latencies of Task Sequences in Multi-Core Automotive ECUs}

\author{\IEEEauthorblockN{Max J. Friese\IEEEauthorrefmark{1}\IEEEauthorrefmark{2},
						  Thorsten Ehlers\IEEEauthorrefmark{1} and 
						  Dirk Nowotka\IEEEauthorrefmark{1}}
	\IEEEauthorblockA{
		\IEEEauthorrefmark{1}
		\textit{Department of Computer Science} \\
		\textit{Kiel University}\\
		Germany \\
		\{mjf,the,dn\}@informatik.uni-kiel.de}  
	\IEEEauthorblockA{
		\IEEEauthorrefmark{2}
		\textit{Mercedes-Benz Car Development} \\
		Sindelfingen, Germany \\
		max\_jonas.friese@daimler.com}
	}

\begin{document}
	
	\maketitle

	\begin{abstract}
		The computation of a cyber-physical system's reaction to a stimulus 
		typically involves the execution of several tasks. 
		The delay between stimulus and reaction thus depends on the interaction of these 
		tasks and is subject to timing constraints. 
		Such constraints exist for a number of reasons and range from possible impacts on 
		customer experiences to safety requirements. 		
		We present a technique to determine end-to-end latencies of such task sequences. 
		The technique is demonstrated on the example of electronic control units (ECUs)
		in automotive embedded real-time systems.
		Our approach is able to deal with multi-core architectures and supports
		four different activation patterns, including interrupts. It is the first
		formal analysis approach making use of load assumptions in order
		to exclude infeasible data propagation paths without the knowledge of 
		worst-case execution times or worst-case response times. 
		We employ a constraint programming solver to compute bounds on end-to-end latencies.  
	\end{abstract}

	\section{Introduction}
	Cyber-physical systems (CPS) are an ubiquitous part in today's 
	connected world. This also spreads to the automotive industry.
	Application areas in this domain range from in-car 
	entertainment systems over driver assistance to engine 
	control. Many features are implemented by electronic control units 
	(ECU) and have real-time requirements which, if not met, can lead 
	to bad customer experience or possibly even safety risks. 
	Consequently timing analysis is an important 
	part of system engineering and it is considerably complex.
	Response times for given stimuli can usually
	be determined in three ways. They can be measured in the end-product or with	
	testbed hardware, determined with simulation or estimated by analytical 
	approaches, as described in \cite{Santos17}. 
	Measurements and simulations have two drawbacks.
	Firstly, they can only be performed relatively late in the engineering process. 
	Secondly, it cannot be assured that the worst case behavior is captured 
	and in-depth knowledge of the system under investigation is necessary.
	Analytical approaches on the other hand	do not suffer from the problem of a
	possibly non-ex\-haust\-ive coverage, but tend to 
	overestimate due to necessary abstractions in the modeling 
	process. Reducing overestimations while not increasing computation 
	time is still becoming more challenging due to the increasing 
	amount of~ECUs and software in modern cars. Reacting to this trend, in order 
	to ensure timing analysis, the development cooperation AUTOSAR extended its 
	standards by a formal timing model as described in \cite{AUTOSAR-TimEx}. 
	These extensions enable AUTOSAR users to give timing 
	restrictions and requirements on different abstraction levels.
	
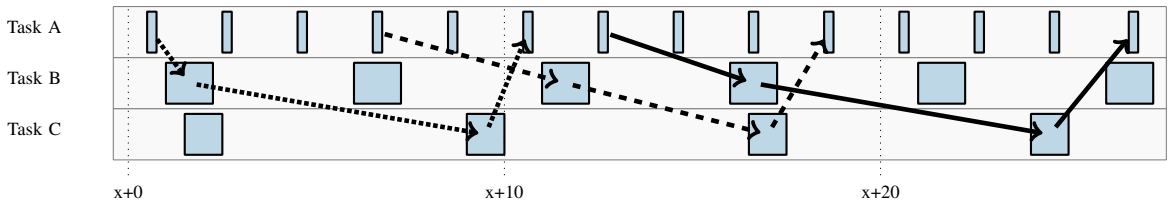
\begin{figure*}[tb]
		\centering
		\vspace*{0.25\baselineskip} \scriptsize
		\begin{tikzpicture}[xscale=0.50,yscale=0.68]
		
		\node[] at (-2.5,-0.5) (1er) {Task A}; 
		\fill[rounded corners=.2pt, draw=gray, fill=gray!5!white]
		(-0.4,-1.1) rectangle ++(28.0, 1.0);
		
		\node[] at (-2.5,-1.5) (1er) {Task B}; 
		\fill[rounded corners=.2pt, draw=gray, fill=gray!5!white]
		(-0.4,-2.1) rectangle ++(28.0, 1.0);
		
		\node[] at (-2.5,-2.5) (1er) {Task C}; 
		\fill[rounded corners=.2pt, draw=gray, fill=gray!5!white]
		(-0.4,-3.1) rectangle ++(28.0, 1.0);

		\foreach \x [count = \xi] in 
		{  0.5,  2.5,  4.5,  6.5,  8.5, 10.5, 12.5, 14.5, 16.5, 18.5, 
			20.5, 22.5, 24.5, 26.6 }
		{	             
			\fill[thick, rounded corners=.2pt, draw=black, fill=fg0light]
			(\x,-1) rectangle ++( 0.25, 0.80) node[pos=0.5] (2er\xi) { }; 
		}	
		
		\foreach \x [count = \xi] in {  1.0,  6.0, 11.0, 16.0, 21.0, 26.0 }
		{	             
			\fill[thick, rounded corners=.2pt, draw=black, fill=fg0light]
			(\x,-2) rectangle ++( 1.25, 0.80) node[pos=0.5] (5er\xi) { }; 
		}	
		
		\foreach \x [count = \xi] in {  1.5,  9.0, 16.5, 24.0 } 
		{	             
			\fill[thick, rounded corners=.2pt, draw=black, fill=fg0light]
			(\x,-3) rectangle ++( 1.00, 0.80) node[pos=0.5] (7er\xi) { }; 
		}

		\foreach \x in {0,10,20} {
			\draw[dotted] (\x,-3.4) -- (\x,-0.1) node[pos=-0.1] (\x)
			{x+\pgfmathparse{int((\x)}\pgfmathresult}; 
		}
		
		\tikzstyle{chain} = [line width=1.75pt, color=black, ->]
		
		
		\draw[chain, densely dotted] (2er1) -- (5er1);
		\draw[chain, densely dotted] (5er1) -- (7er2);
		\draw[chain, densely dotted] (7er2) -- (2er6);
		
		\draw[chain, dashed]         (2er4) -- (5er3);
		\draw[chain, dashed]         (5er3) -- (7er3);
		\draw[chain, dashed]         (7er3) -- (2er10);
		
		\draw[chain, solid]          (2er7) -- (5er4);
		\draw[chain, solid]          (5er4) -- (7er4);
		\draw[chain, solid]          (7er4) -- (2er14);
		
		\end{tikzpicture}
		\caption{Example of a task-level chain with possible instance-level flows}
		\label{fig:chainExample}
	\end{figure*}

	The analysis of software timings has consequently been subject to a wide 
	range of scientific research, e.g.~\cite{TindellClark94,Henia05}. 
	Here, response time is defined as the time between activation 
	and completion of a single task. In current cyber-physical systems 
	the data flow which is needed to provide a function usually 
	passes through multiple task instances possibly being executed on different 
	processing units. These tasks can be activated periodically 
	with different rates, as depicted in~\autoref{fig:chainExample}. The 
	varying relative offsets of the tasks cause different end-to-end
	timings. In this rather small example there are three different 
	instantiations for one sequence of tasks. 
	
	There are different kinds of temporal constraints on such task sequences. 
	The delay between the arrival of an input and the reaction of the last task 
	of the task sequence is called \emph{response time}. 
	On the contrary, \emph{data age} describes the 
	fact that outputs may depend on old input values, and there may be some delay 
	before they are updated to reflect later inputs. 
	The differentiation between these concepts was firstly considered 
	in~\cite{Feiertag2008}. 
	The data-dependency between two task instances is 
	also referred to as \emph{job-level dependencies} 
	(cf.\cite{DBLP:conf/rtcsa/BeckerDMBN16}). 
	Task sequences with such dependencies are also called 
	\emph{cause effect chains}. Closely related to them 
	are so-called \emph{event chains} where the system behavior is 
	described in terms of event models as in \cite{DBLP:conf/rtas/SchlatowE16}. 
	Regardless of the type of constraints, it is preferable to 
	have guarantees about them in early development stages. Therefore 
	the use of formal methods using abstract models, e.g. 
	constraint programming, is promising. 
	Here, we focus on the response time in cause effect chains. 

	This paper presents an approach for the formal analysis of 
	task chains. The present approach does not require an estimation for the
	actual execution time of the executables of a task and is therefore 
	applicable in early development stages.
		
	We use an intuitive task model which needs very few information 
	about the task set but gives the possibility to add more details
	at a later stage.
	The communication between task instances is assumed to happen 
	via signals which are located in a shared memory. We distinguish 
	between three communication models. In explicit communication values are read and written 
	immediately. Implicit communication means that each task instance 
	creates a local copy of all signals and writes them immediately 
	after termination. In the case of deterministic communication,  
	a task instance only writes at  
	well-defined points of time relative to its activation.
	In our approach, a system is encoded in a constraint 
	program and a solver is deployed to calculate the 
	worst case response time for a task sequence. Other approaches 
	to complete this task include the compositional (or modular) analysis, 
	exhaustive search and model checking e.g. using hybrid automata. 
	Related work will be presented in the next section. 
	Subsequently, we give a more detailed description
	of the system that we model and spell out the premises for our approach. 
	Based on this, we define a set of constraints according to our assumptions.
	Finally, we discuss the results of the analysis of some realistic examples 
	from an automotive supplier and a manufacturer of automobiles.

	\section{Related Work}
	Work most related to ours comes from two fields: 
	modeling systems and describing timing behavior, and the estimation of end-to-end 
	latencies. In the automotive domain, a popular representative of the former
	is AUTOSAR. The AUTOSAR standards' timing model 
	is based on the Timing Augmented Description Language (TADL2)
	which was developed in the Timing Model (TIMMO, \cite{TimmoWeb}) 
	project with the goal to
	standardize timing descriptions in automotive real-time systems 
	as cited in \cite{TimmoProjectWeb}. Another model language which is 
	focused on timing is 
	the \emph{time-triggered language for embedded programming} Giotto 
	which was introduced in \cite{Henzinger2001} and is the 
	basis for the logical execution time (LET) paradigm. 
	LET considers abstract
	intervals between the reading and writing of variables 
	instead of the actual execution time of a program and is focused on
	software only. In the automotive domain a case of application is 
	e.g. the distribution of single-core software on multi-core platforms 
	\cite{DBLP:conf/rtas/HennigHMRLN16}. It is also a possible basis for  
	new approaches to increase timing predictability of embedded
	real-time systems \cite{Henzinger2001,Kirsch2012}. 
	
	The second category of related work is about performance analysis with regards 
	to end-to-end timing. Different understandings of \emph{timing} exists and 
	there is a subtle difference in notion. 
	On one hand, end-to-end delays
	refer to the time  
	data propagation needs to take place on a specific path in a system were tasks 
	are triggered independently \cite{DBLP:journals/jsa/BeckerDMBN17}. 
	On the other hand, response times refer to the response of the last 
	task in a chain of tasks where task executions are triggered by events, 
	including a triggering from another task \cite{DBLP:conf/ecrts/GutierrezGH97}. 
	Furthermore, in the context of cyber-physical systems the term output latency 
	is used for the time between stimulus and response of the system 
	\cite{DBLP:conf/codes/ShrivastavaDLSK16}. 
	We will stick to the term output latency or latency for short. 
	Regardless of the notation, analysis gets harder when multi-core systems get involved.  
    An approach for the formal estimation of latencies
    for multi-rate
    cause effect chains supporting multi-core systems was presented 
    in \cite{Feiertag2008}. Here multi-rate means that the chain contains 
    tasks with different periods.     
    It supports different path semantics yielding different 
    time constraints and is based on an event model of the system. 
	In \cite{DBLP:conf/rtcsa/BeckerDMBN16} Becker et al. introduce the
	notion of job-level dependencies. Building on that 	
	they show in \cite{DBLP:journals/jsa/BeckerDMBN17} how this 
	approach can be used to determine end-to-end data age at different 
	levels of knowledge about the system. The safe estimations however 
	are rather pessimistic when compared to end-to-end data ages 
	determined with given schedules. 
	
	Prior to the use of multi-core systems, response time analysis was focused 
	on single tasks, e.g. \cite{Tindell94}. 
	Based on event models, Richter and Ernst presented a compositional method in  
	\cite{DBLP:conf/dac/RichterZJE02} which can be used to connect different 
	analyses via input and output models. It combines different local response time 
	calculations to model the global system behavior. This idea has also been 
	adapted for multi-core systems. 
	For chains which are formed by tasks which trigger their respective successor, 
	a compositional approach based on event models of the system 
	was presented in \cite{DBLP:conf/rtas/SchlatowE16}. 
	A range of industrial tools for the analysis of heterogeneous multiprocessor systems 
	exists~\cite{DBLP:conf/icst/RiouxHS17,ArcticusWeb,LuxoftWeb,TimingArchitectsWeb}.
	Some of them can also be used to perform timing analysis on multi-rate 
	cause effect chains. These tools however rely on information which is only
	available at implementation level, e.g. execution time measurements. 
	
    The aforementioned approaches follow an algorithmic approach, 
 	e.g. using backtracking. Another approach to estimating 
 	end-to-end latencies is based on mixed integer linear programming (MILP) 
 	and was presented in \cite{DBLP:conf/nfm/BoniolLPE13,DBLP:journals/ijccbs/LauerBPE14}.
 	There, the system's behavior and the timing
 	properties of \emph{functional chains} are modeled in terms of linear constraints. 
	However, the setting in these papers differs from the one considered here. 
	Firstly, another scheduling model is used. Secondly, the possible interferences of tasks are assumed to be 
	covered in an earlier analysis step, as it makes use of the results 
	of a worst-case response time analysis. 
	Furthermore, the results from these papers suggest that the non-convexity of the problem 
	makes it hard for MILP solvers. 
	
 	%
 	To the best of our knowledge no descriptive approach using a 
 	constraint modeling language has been presented for the problem of 
 	estimating end-to-end latencies for multi-rate cause effect 
 	chains. 	
 	Our approach makes use of a set of 
    constraints for modeling the system and the task sequence for the chain. 
	Thereupon we deploy a constraint solver for estimating latencies.  
 	The latencies are safe in the sense that no worse latency is 
 	observable in the real system, assuming that no overload occurs. 
 	 	
	\section{Task Sequences as Constraint Program}
	The presented approach is based on an a declarative description of all 
	possible schedules via a set of constraints.
	Searching for the worst case end-to-end latency in the satisfying
	assignments for these constraints simulates an exhaustive 
	search. The end-to-end latency of a sequence of tasks 
	for the scope of this work then is the difference in time 
	between the activation of the first task on the chain and 
	the point of time where the last task on the chain wrote 
	its results. This sequential analysis also suffices for the 
	end-to-end timing analysis of cause-effect chains \cite{DBLP:journals/jsa/BeckerDMBN17}. 
	
	The descriptive nature of this approach allows for adapting the
	model, e.g. with respect to new constraints, without the need of adjusting 
	possibly complex algorithms. Furthermore we can choose from a wide range of 
	solver-backends, some of which support parallel computing.
	 	
	The contributions of this work are twofold. 
	Firstly, we 
	introduce the constraints needed for a description of a task set in an 
	early stage of the engineering process. Secondly, we show how a constraint solver 
	can be utilized to make use of new formal approaches to end-to-end timing 
	estimations. \todo{removed redundant part}
	

	In the following we present the system to be modeled and the 
	task chain definition we use. In the main part of this section we 
	are concerned with the formal constraints and the interval we need to 
	consider for safe estimations. 
	
	\subsection{System Model}
	The task set of ECUs in automotive real-time systems commonly consists 
	of two different types of tasks. On the one hand there are periodic tasks 
	which may have an offset. On the other hand, event-based tasks 
	occur sporadically.  	
	Instances of tasks communicate to propagate information on   
	specific paths in order to implement system functions. 
	We consider multiprocessor systems deploying a finite set of 
	tasks which communicate via variables which are located in 
	shared memory. 
	We focus on fixed priority preemptive scheduling (FPPS) in this work. 
	Other scheduling policies can be implemented but possibly need further 
	variables and constraints. 
	
	Due to our focus on early development stages, we make some assumptions. 
	Firstly, we assume that every task has a minimal execution time, which may equal zero. 
	Secondly, we assume that every task execution finishes before the respective deadline. 
	This appears reasonable as a system should not be overloaded by design. 
	Furthermore, task interferences by e.g. critical sections are handled implicitly, assuming 
	that tasks do not miss their deadline no matter which interferences occur. 
	
    \todo{new par}
		
	Two basic communication paradigms are relevant 
	in the automotive domain: implicit and explicit communication~\cite{AUTOSAR-RTE}. 
    When implementing explicit communication, every task reads and writes
	the data it processes immediately from and to shared memory. 
	In implicit communication, each task 
	instance creates a local copy of all data it processes. This data 
	is manipulated while the execution of the task and written back to the 
	global memory after its termination. 
	
	We consider a third paradigm 
	which is a variation of the implicit communication as this is only 
	applicable for tasks which are periodically activated. 
	After the termination of a task instance all data is hold 
	in a local buffer and written 
	back only at well-defined points of time which are relative to the 
	activation of the task instance, i.e. before the next instance of the 
	task is activated. We will also refer to this as \emph{deterministic} communication. 
	\autoref{fig:detTiming} and~\autoref{fig:nondetTiming} show the differences
	in communication for two tasks with different activation intervals. 
	In~\autoref{fig:detTiming} Task $A$ is only allowed to propagate its results 
	at points of time marked with a dashed line (e.g. at $t=x+10$).
	Task $B$ is only allowed to do so at points of time marked with a dotted or 
	a dashed line (e.g. at $t=x+5$ or at $t=x+10$). 
	The dashed arrows show reads and writes which are not possible here but would 
	be possible in the non-deterministic case.	
	Due to these restrictions, the second instance of Task B cannot use the data 
	written by the first instances of Task A, and operates on old data. 
	
	\begin{figure}[h]
		\centering
		\begin{tikzpicture}[xscale=0.33,yscale=0.42] \scriptsize
		
		\fill[thick, rounded corners=.2pt, draw=black, fill=gray!50!white]
		(-0.4, 3.6) rectangle ++(20.8, 1.0)
		node[pos=0.5] (RAM) {Shared Memory};
		
		\node[] (taskA) at (-1.5, 2.0) { Task $A$ }; 
		\node[] (taskB) at (-1.5, 0.5) { Task $B$ };
		
		\draw [draw=none, fill=white!40!gray, opacity=0.20]
		( 0.0, 0.0) -- (20.0, 0.0) -- (20.0, 1.0) -- ( 0.0, 1.0) -- cycle;
		
		\draw [draw=none, fill=white!40!gray, opacity=0.20]
		( 0.0, 1.5) -- (20.0, 1.5) -- (20.0, 2.5) -- ( 0.0, 2.5) -- cycle;

		\foreach \x in {5,15} {
			\draw[dotted] (\x,3.0) -- (\x,-0.5) node[below right = .1 and -.1] (\x)
			{x+\pgfmathparse{int((\x)}\pgfmathresult}; 
		}
		\foreach \x in {0,10,20} {
			\draw[thick,dashdotted] (\x,3.0) -- (\x,-0.5) node[below right = .1 and -.1] (\x)
			{x+\pgfmathparse{int((\x)}\pgfmathresult}; 
		}
		
		\draw[->] (-0.5, -0.5) -- (22.0, -0.5);
		
		
		\foreach \x[evaluate={\n=int(\x*0.2)}] in {2.5,7.5,12.5,17.5} {
			\fill[thick, rounded corners=1pt, draw=black, fill=fg1]
			( \x, 0.1) rectangle ++( 1.8, 0.8)
			node[pos=0.5] (5er\n) {}; 
		}
		
		\foreach \x[evaluate={\n=int(\x*0.1)}] in {0,10} { 
			\fill[thick, rounded corners=1pt, draw=black, fill=fg0]
			( \x, 1.6) rectangle ++( 2.5, 0.8)
			node[pos=0.5] (10erS\n) {}; 
			
			\fill[thick, rounded corners=1pt, draw=black, pattern=north east lines, pattern color=fg0]
			(2.5+\x, 1.6) rectangle ++( 1.8, 0.8);
			
			\fill[thick, rounded corners=1pt, draw=black, fill=fg0]
			(4.3+\x, 1.6) rectangle ++( 2.3, 0.8)
			node[pos=0.5] (10erE\n) {}; 
		}
		
		\coordinate[right = 0.2 of 10erE0]  (p0) ;
		\coordinate[right = 0.1 of   5er2]  (p1) ;
		\coordinate[left  = 0.1 of   5er2]  (p2) ;
		\coordinate[left  = 0.6 of 10erS1]  (p3) ;
		\coordinate[left  = 0.1 of   5er1]  (p1M) ;

		\draw[line width=1.80pt, color=black, - ] (p0) -- (p3);
		\draw[line width=1.80pt, color=black, ->] (p3) --+ ( 0.0, 1.9);
		\draw[line width=1.80pt, color=black, ->] (p1) --+ ( 0.0, 3.4);
		\draw[line width=1.80pt, color=black, <-] (p2) --+ ( 0.0, 3.4);

		\draw[line width=1.80pt, color=black, dotted, ->] (p0) --+ ( 0.0, 1.9);
		\draw[line width=1.80pt, color=black, dotted, <-] (p1M) --+ ( 0.0, 3.4);
		
		\end{tikzpicture}
		\caption{Task communication in deterministic case}
		\label{fig:detTiming}
	\end{figure}
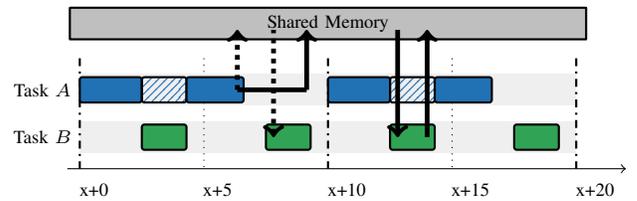	
	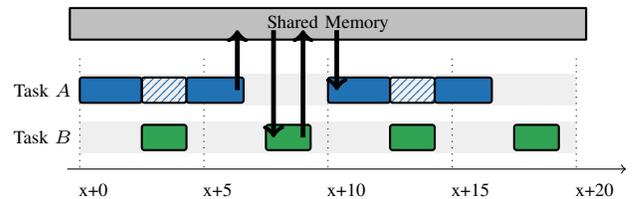
\begin{figure}[h]
		\centering
		\begin{tikzpicture}[xscale=0.33,yscale=0.42] \scriptsize
		
		\fill[thick, rounded corners=.2pt, draw=black, fill=gray!50!white]
		(-0.4, 3.6) rectangle ++(20.8, 1.0)
		node[pos=0.5] (RAM) {Shared Memory};
		
		\node[] (taskA) at (-1.5, 2.0) { Task $A$ }; 
		\node[] (taskB) at (-1.5, 0.5) { Task $B$ };
		
		\draw [draw=none, fill=white!40!gray, opacity=0.20]
		( 0.0, 0.0) -- (20.0, 0.0) -- (20.0, 1.0) -- ( 0.0, 1.0) -- cycle;
		
		\draw [draw=none, fill=white!40!gray, opacity=0.20]
		( 0.0, 1.5) -- (20.0, 1.5) -- (20.0, 2.5) -- ( 0.0, 2.5) -- cycle;

		\foreach \x in {0,5,15} {
			\draw[dotted] (\x,3.0) -- (\x,-0.5) node[below right = .1 and -.1] (\x)
			{x+\pgfmathparse{int((\x)}\pgfmathresult}; 
		}
		\foreach \x in {10,20} {
			\draw[dotted] (\x,3.0) -- (\x,-0.5) node[below right = .1 and -.1] (\x)
			{x+\pgfmathparse{int((\x)}\pgfmathresult}; 
		}
		
		\draw[->] (-0.5, -0.5) -- (22.0, -0.5);
		
		
		\foreach \x[evaluate={\n=int(\x*0.2)}] in {2.5,7.5,12.5,17.5} {
			\fill[thick, rounded corners=1pt, draw=black, fill=fg1]
			( \x, 0.1) rectangle ++( 1.8, 0.8)
			node[pos=0.5] (5er\n) {}; 
		}
		
		\foreach \x[evaluate={\n=int(\x*0.1)}] in {0,10} { 
			\fill[thick, rounded corners=1pt, draw=black, fill=fg0]
			( \x, 1.6) rectangle ++( 2.5, 0.8)
			node[pos=0.5] (10erS\n) {}; 
			
			\fill[thick, rounded corners=1pt, draw=black, pattern=north east lines, pattern color=fg0]
			(2.5+\x, 1.6) rectangle ++( 1.8, 0.8);
			
			\fill[thick, rounded corners=1pt, draw=black, fill=fg0]
			(4.3+\x, 1.6) rectangle ++( 2.3, 0.8)
			node[pos=0.5] (10erE\n) {}; 
		}
		
		\coordinate[right = 0.2 of 10erE0]  (p0) ;
		\coordinate[left  = 0.1 of   5er1]  (p1) ;
		\coordinate[right = 0.1 of   5er1]  (p2) ;
		\coordinate[left  = 0.2 of 10erS1]  (p3) ;
		
		\draw[line width=1.80pt, color=black, ->] (p0) --+ ( 0.0, 1.9);
		\draw[line width=1.80pt, color=black, <-] (p1) --+ ( 0.0, 3.4);
		\draw[line width=1.80pt, color=black, ->] (p2) --+ ( 0.0, 3.4);
		\draw[line width=1.80pt, color=black, <-] (p3) --+ ( 0.0, 1.9);
		
		\end{tikzpicture}
		\caption{Task communication in non-deterministic case}
		\label{fig:nondetTiming}
	\end{figure}	
	In contrary, Figure~\ref{fig:nondetTiming} shows non-deterministic communication. 
	Here data processed by Task A can be read by Task B immediately

	For the analysis of the end-to-end timing of a sequence of  
	tasks it is furthermore necessary to consider all possible relative offsets of instances 
	of these tasks. The relative offsets vary over time with the  
	observable schedules. The amount of possible schedules again depends 
	on several properties of the tasks, the scheduler, and the mapping of tasks to cores. 
	These key properties hence form the base for the system's encoding as 
	a set of constraints. For a task at least an activation pattern 
	and a deadline are needed. Given this information and a scheduling policy we can 
	represent a superset of all possible schedules as the set of solutions of our constraint program. 
	The analysis will yield 
	more precise results if more details of the system are encoded in 
	the constraints as described below.
	\todo{new par. Lass ma am Telefon darueber quatschen}
	Within the superset of possible schedules we then search for all data propagation
	paths between the instances of the task sequence. 
	
	Additional details of the system can be used to add more constraints and prune possible schedules. 
	This reduces the set of feasible propagation paths of the task sequence, thus increasing precision. 
	A natural refinement is adding details about the actual core execution times
    of the task instance and the resulting scheduling decisions. In the following 
    we discuss several ways to decrease the amount of considered schedules and 
    therefore data propagation paths by refining the system model. 
    
    \subsection{Model Refinements}
    The first refinement is that a task instance needs a
    minimum amount of resources to be processed, i.e. a best case execution time (BCET). 
    Consequently its results are not 
    available before some point of time relative to its start. 
    This can result in the infeasibility of some propagation paths, namely the ones 
    where a task cannot propagate its results because it could not have possibly produced it. 
    
    Secondly, a task instance might block other tasks or cause the scheduler to 
    pause them due to higher priority. 
    These interferences again might result in the infeasibility of some propagation 
    paths, e.g. when a low-priority task is always delayed and can not possibly propagate 
    its results to the, in terms of activation time, nearest instance of a higher priority task.
    A benefit of the descriptive approach is that we are able to adjust 
	the level of detail which is provided to the solver depending on the amount of information 
	available in the respective development stage. 
	For example, another refinement 
	which is not considered for now due to our focus on early design stages, 
	is concerned with the software modules of a task. When a task is activated 
	some executable code is processed which typically is assembled from 
	one or more software modules. This is why a promising refinement 
	is adding the exact order of the software modules in each task with best-
	and worst-case execution times. 

    
    \subsection{Activation Patterns}
    An activation pattern describes the temporal nature of how the instances 
    of a task occur. 
    	
    The first activation pattern is called periodic activation with offset. 
	It describes a pattern which is widely used in embedded systems. 
	A task is activated after a specific and fixed amount of time has passed. 
	We also consider the extended variant with offsets, meaning that the first 
	instance of this task is not activated at time zero but time zero plus 
	offset. 	
	
	The second activation pattern we consider is \emph{chained activation}. 
	A task which is chained to an other task is activated when its predecessor 
	triggers its execution. This activation can be used to induce controlled 
	concurrency. Assume a task A fulfils some function on a 
	processing unit which should not interfere with the execution of 
	another task B on a second processing unit. To ensure this, 
	task B can be chained to task A. Thus, task B is activated when task A finishes
	its execution, which prevents them from running at the same time. 
     	
	The third activation pattern is taken into consideration to handle 
	angle synchronous tasks. 
	Many time-critical functions in the automotive domain are related with the 
	gearbox and the engine. The engine control unit e.g. computes the right ignition timing. 
	This needs to be done 
	synchronously to the engine speed. The timing properties of these 
	tasks can hardly be modeled with an offset and period without adding too much pessimism.
	The nature of these activation patterns is rather a minimum and 
	maximum occurrence per time interval, e.g. because the rounds per minute of the engine
	cannot be arbitrarily high. This can be converted 
	to a minimum and maximum temporal distance of two consecutive 
	activation events as shown in \cite{DBLP:conf/codes/SchlieckerE09}. 
	We call this pattern the \emph{bounded activation}.		
	
	The fourth activation pattern is needed to describe tasks which 
	are activated sporadically with less information about the 
	relative offset between two consecutive activations. These are for 
	example communication interrupts or sensor readings. In these 
	cases the assumption about a specific pattern needs to be 
	relaxed \cite{DBLP:conf/ecrts/MoyoNLM10}.
	Since these tasks also require a minimum amount of time to be processed, 
	we need an assumption on the minimum of time between two
	consecutive occurrences in order to yield a feasible scheduling. 
	This is a valid assumption since usually there is a debouncing 
	time for sensor readings and a maximal sending rate for network messages. 
	Since we only know the minimum time between two task activations
	but not the maximum time, the theoretical estimation of each chain
	involving one of these tasks is always infinite. We therefore exclude 
	chains consisting of tasks with this activation pattern from our 
	considerations, except when it is located at the very beginning of the chain.
	We then add the assumption that the activation happens within a critical 
	interval which will be discussed below. 
	
	\begin{table}[t]
	\centering
	\caption{Task variables with domain}
	\begin{tabular}{|l p{4.7cm} c|} \hline
		\textbf{Variable}& \textbf{Description} & \textbf{Domain}   \\ \hline
		$\tPeriod$       &%
							Period of a periodically activated task & $\mathbb{N}$ \\
		$\tOffset$       &%
							Offset of a periodically activated task & $\mathbb{N}$ \\
		$\tPredecessor$  &%
							Pointer to the predecessor of a chained task & $\mathcal{T}$ \\
		$\tMinDeltaT$    &%
		                    Minimum temporal distance of two consecutive activations of a sporadic 
		                    task (bounded or interrupt)    & $\mathbb{N}$ \\
		$\tMaxDeltaT$    &%
							Maximum temporal distance of two consecutive activations 
							of a sporadic task with bounded occurrence   & $\mathbb{N}$ \\ \hline
		$\tPrio$         &%
							The priority of the task for fixed priority scheduling & $\mathbb{N}$ \\
		$\tCore$         &%
							The index of the processing unit on which the task is scheduled 
							(we assume an enumeration.) & $\mathbb{N}$ \\
		$\tPree$         &%
							A flag to define whether a task is interruptible & $\set{\mathtt{t},\mathtt{f}}$ \\
		$\tDead$         &%
							The deadline of the task & $\mathbb{N}$ \\
		$\tActT$         &%
							The activation type of the task & $\mathcal{A}$ \\\hline                   
	\end{tabular}
	\label{tbl:TaskVars}
	\end{table}		
	
	The upper half of~\autoref{tbl:TaskVars} shows the variables we use in the 
    constraints for the activation patterns. Let $\mathcal{T}$  be the set of tasks.
	The lower half of~\autoref{tbl:TaskVars} shows the variables needed to describe 
	the task set with the proposed level of detail. We assume that the given core execution 
	times refer to the core given in $\tCore(i)$ f.a. $i\in \mathcal{T}$.
	Let $\mathcal{A} = \set{\mathtt{periodic},\mathtt{chained},\mathtt{bound},\mathtt{sporadic}}$ 
	the set of possible activation types.
	
	\subsection{Problem Encoding}
	
	Until here we named all information we need to describe the task set on the 
	assumed level of detail. In the following, this information is used to describe
	the behavior of the task set when instantiated. This leads to a set of schedules 
	and data flows as described above. 	
	The following constraints must be satisfied:
	\begin{enumerate}
		\item Tasks are activated according to their activation pattern 
		as described above 
		(cf. Constraints~(\ref{eq:alpha0}), (\ref{eq:alpha1}), 
		(\ref{eq:boundFirst0}), (\ref{eq:boundFirst1}), (\ref{eq:dtmin}), 
		and (\ref{eq:dtmax})). 
		\item A task instance is enqueued for scheduling after its
		activation. The task which is scheduled next is selected 
		by the scheduling policy for each processing unit. (cf. Constraint~(\ref{eq:sigma}))          
		\item The processing of a task instance 
		is delayed if another task instance  is processed on the assigned processing unit 
		which is not preemptable or preferred for execution by the scheduling 
		policy (cf. Constraint~(\ref{eq:sigma})). 
		\item A task instance which gets activated may interrupt a currently 
		processed task instance if the latter is preemptable 
		and the former is preferred for execution by the scheduling 
		policy and both are assigned to the same processing unit
		(cf. Constraints~(\ref{eq:iotaHelper}) and (\ref{eq:iota})).
		\item Tasks read and write shared variables according to 
		their communication paradigm which is a fixed one from the 
		above-mentioned (cf. Constraints~(\ref{eq:implicitCommunication}), 
		(\ref{eq:explicitCommunication}), and (\ref{eq:detCommunication})).
	\end{enumerate}
	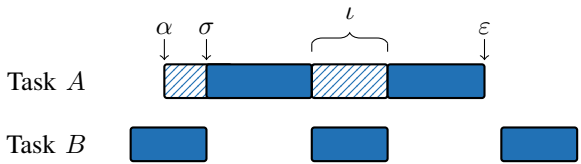
\begin{figure}[h]
		\centering
		\begin{tikzpicture}[xscale=0.56,yscale=0.56] 
		
		\node[] (taskA) at (-3.8, 2.0) { \normalsize Task $A$ }; 
		\node[] (taskB) at (-3.8, 0.5) { \normalsize Task $B$ };
		
		\foreach \x[evaluate={\n=int(\x*0.2)}] in {-1.8, 2.5, 7.0} {
			\fill[thick, rounded corners=1pt, draw=black, fill=fg0]
			( \x, 0.1) rectangle ++( 1.8, 0.8);	
		}
	
		\fill[thick, rounded corners=1pt, draw=black, pattern=north east lines, 
		pattern color=fg0]
		(-1.0, 1.6) rectangle ++( 1.6, 0.8);
		
		\fill[thick, rounded corners=1pt, draw=black, fill=fg0]
		( 0.0, 1.6) rectangle ++( 2.5, 0.8);	
		
		\fill[thick, rounded corners=1pt, draw=black, pattern=north east lines, 
		pattern color=fg0]
		( 2.5, 1.6) rectangle ++( 1.8, 0.8);
		
		\fill[thick, rounded corners=1pt, draw=black, fill=fg0]
		( 4.3, 1.6) rectangle ++( 2.3, 0.8);	
		
		\draw[<-] (-1.0, 2.5) -- (-1.0, 3.0) node[pos=1.5]  {$\alpha$}; 
		\draw[<-] ( 0.0, 2.5) -- ( 0.0, 3.0) node[pos=1.5]  {$\sigma$}; 
		\draw[ -] ( 2.5, 2.7) -- ( 2.5, 2.9) ;  
		\draw[ -] ( 4.3, 2.7) -- ( 4.3, 2.9) ;
		\draw [decorate,decoration={brace,amplitude=4pt}] 
		( 2.5, 2.9) -- ( 4.3, 2.9) node[midway, above=.2cm] {$\iota$};
		\draw[<-] ( 6.6, 2.5) -- ( 6.6, 3.0) node[pos=1.5]  {$\varepsilon$};
		
		\end{tikzpicture}
		\caption{Variables of a task instance}
		\label{fig:taskVars}
	\end{figure}

	In order to model the constraints 1) -- 6) each
	task instance is described with four variables: 
	an activation time $\alpha$, 
	an interrupted time $\iota$, 
	an actual starting time $\sigma$ and 
	a finishing time $\varepsilon$. 
    \autoref{fig:taskVars} shows the 
	variables in the context of the lifespan of a task instance. Task $B$ in 
	this figure is a task which delays and interrupts task $A$ for illustrative 
	purposes. Task instances may interact in two ways. Firstly one task instance 
	can delay the execution of instances of other tasks. This is expressed 
	in the differences between $\alpha$ and $\sigma$. Secondly 
	one task instance may pause the execution of instances of 
	other tasks. This is expressed in the $\iota$ of the paused 
	instance. This variable contains the total amount of time for which a task instance  
	was paused after its start. 
	
	In the following, we give constraints which must hold for a schedule which starts at 
	time $0$, and where a task $i \in \mathcal{T}$ occurs $m_i$ times. 
	The choice of the values $m_i$ will be discussed in detail in Subsection~\ref{sec:relevantPeriod}.

  	The constraints on the $\alpha$s depend on the activation pattern 
  	of the respective task. Periodic tasks are activated at multiples of their period, starting at their offset.
  	\begin{align}\label{eq:alpha0}
  		\alpha_{i,j} = \tOffset(i) + j \cdot \tPeriod(i) && \forall j\colon 1 \leq j \leq m_i
  	\end{align}
  	In the case of chained activation, an instance of task $i$ 
  	is activated if its predecessor $p(i)$ is terminated. This is, 
  	\begin{align}\label{eq:alpha1}
  		\alpha_{i,j} &= \varepsilon_{p(i),j} && \forall j\colon 1 \leq j \leq m_i
  	\end{align}
  	for each task $i\in\mathcal{T}$ where $\tActT(i)=\mathtt{chained}$. 
  	
  	For a task $i\in\mathcal{T}$ which is activated non-periodically according
  	to a $\mathtt{sporadic}$ activation pattern, the time between two activations 
  	is in the interval $[\tMinDeltaT(i), \tMaxDeltaT(i)]$. 
  	Thus, for a task $i$ where $\tActT(i)=\mathtt{bounded}$ the activation time is constrained by 
  	\begin{align}
  		\alpha_{i,1} &\geq 0 			       \label{eq:boundFirst0} \\
  		\alpha_{i,1} &\leq \tMaxDeltaT(i)      \label{eq:boundFirst1} \\
  		\alpha_{i,j} &\geq \alpha_{i,j-1} + \tMinDeltaT(i)  && \forall j\colon 2 \leq j \leq m_i\label{eq:dtmin} \\
  		\alpha_{i,j} &\leq \alpha_{i,j-1} + \tMaxDeltaT(i)  && \forall j\colon 2 \leq j \leq m_i\label{eq:dtmax}
  	\end{align}
  
  	The constraints~\ref{eq:boundFirst0} and~\ref{eq:boundFirst1}
  	assure the consideration of all possible relative offsets 
  	with other tasks which we need to ensure a safe estimation. 
  	
  	Lastly, for tasks which occur in a sporadic manner the same constraints 
  	except the one in~\autoref{eq:dtmax} and~\autoref{eq:boundFirst1} hold. This 
  	means that the first instance might occur in $[0,{t^{\text{P}_{\text{min}}}}]$ where 
  	$\tActT(i)=\mathtt{sporadic}$. 

	After a task was activated, some time may pass before it is actually started.
	The following constraints correspond to FPPS, 
  	other policies can be implemented in a very similar fashion. In case of FPPS 
  	a task instance is delayed when another task instance is currently being processed on the 
  	same processing unit which either has a higher priority or is not preemptable.  
  	Thus, the time at which a task instance is actually started may either be its activation time or the 
  	time at which another task finishes its execution.  
  	We therefore consider two sets of termination times with possible impact on the 
  	delay of a task instance. 
  	For tasks with higher priority, we consider all termination times of task instances which 
  	are activated before 
  	$\sigma_{i, j}$, as these will be processed first. 
  	This set is denoted  $D^{\text{HP}}_{i,j}$ and defined as:  
  	\begin{align*}
  	D^{\text{HP}}_{i,j} = \bigl\{\varepsilon_{\ell,k}  \bigm|\;&\alpha_{\ell,k} \leq \sigma_{i, j}\land     \\
  	&\tCore(i) = \tCore(\ell) \land                                              \\
  	&\tPrio(i) < \tPrio(\ell) \bigr\} .                       
  	\end{align*}
  	
  	For non-preemptive tasks we consider the set of task instances which are started 
  	before the $j$-th instance of task $i$ is activated. 
  	These will not be interrupted by the scheduler, and may therefore delay the start of task $i$. 
  	Their execution times are denoted $D^{\text{NP}}_{i,j}$ for  $i,\ell\in \mathcal{T}$, $ 1 \leq j \leq m_i$ 
  	and $1 \leq k\leq  m_\ell$: 
  	\begin{align*}                                                                                  
  	D^{\text{NP}}_{i,j} = \bigl\{ \varepsilon_{\ell,k} \bigm|\;&\sigma_{\ell,k} \leq \alpha_{i, j}\land   \\
  	                     &\tCore(i) = \tCore(\ell) \land                                            \\
  	                     &\tPree(\ell) = \texttt{false} \bigr\} .                                           
  	\end{align*}                                                                                    
  	Then the constraints regarding the actual start of a task instance are 
  	f.a. $i\in \mathcal{T}$ and $1 \leq j \leq m_i$:
 	\begin{align}\label{eq:sigma}
  	\sigma_{i,j} = \max\left(D^{\text{HP}}_{i,j} \cup D^{\text{NP}}_{i,j} \cup \set{\alpha_{i,j}}\right) . 
  	\end{align}
  	While running, a task instance  might be interrupted by other task instances 
  	which are scheduled to the same processing unit. The current 
  	instruction pointer is stored and, in the case of FPPS, the 
  	higher priority task is processed. After the return of the 
  	latter task, the instruction pointer is restored and the remaining 
  	instructions of the interrupted task instance are processed, if no further 
  	interrupt occurs. To model this behavior we introduce a variable 
  	to sum up the time an instance of a task was paused. 
  	With FPPS, a task instance might 
  	have such paused times when the corresponding task is 
  	preemtable and a higher priority task which is mapped to the same 
  	processing unit is activated during its lifetime. 
  	We therefore define f.a. $i,\ell \in \mathcal{T}$, $j\in \set{1,\dots,m_i}$ 
  	and $k\in  \set{1,\dots,m_\ell} $ the following function: 
  	\begin{align} \label{eq:iotaHelper}
  	\begin{split}
  	i^{\text{HP}}(i,j,\ell,k) = 
  	\begin{cases}
  	1 & \text{ if } \sigma_{\ell, k} > \sigma_{i, j} \wedge \varepsilon_{\ell, k} < \varepsilon_{i, j} \land \\
  	  & \quad       \tCore(i) = \tCore(\ell) \land  \\
  	  & \quad       \tPrio(i) < \tPrio(\ell)        \\
  	0 & \text{ otherwise } 
  	\end{cases}
  	\end{split} . 
  	\end{align}
  	This is, $i^{\text{HP}}(i,j,\ell,k)$ returns $1$ if the $j$-th 
  	instance of task $i$ was paused by the scheduler because of the 
  	$k$-th instance of task $\ell$. 
  	Thus the constraints in~\autoref{eq:iota} model the paused time
  	for each task instance. The task instance causing the pause 
  	is guaranteed to finish before the paused instance $j$ of task $i$. 
  	We need to subtract the time the former task instance was paused itself. 
    For all $i\in \mathcal{T}$ and $1 \leq j \leq m_i$  we thus define the following 
    constraint: 
  	\begin{align}\label{eq:iota}
  	\iota_{i,j} = \sum_{\ell\in \mathcal{T} \setminus \{i\}} 
  	\left( 
  		\sum_{k = 1}^{m_\ell} 
  		i^{\text{HP}}(i,j,\ell,k) \cdot 
  		\left( 
  		\varepsilon_{\ell,k} - \sigma_{\ell,k} - \iota_{\ell,k} 
  		\right)
  	\right) . 
  	\end{align} 
  	
  	The last variable describing the lifetime of a task instance is $\varepsilon$ which 
  	models the point of time when all software modules of the task instance 
  	have been processed. This point of time depends on two factors: the actual starting time of the instance, and it depends on how long the actual code execution for the 
  	instance takes. Since we a assume an early stage of development, we only 
  	assume a best-case execution time for the modules. 
  	
  	This is, we can give an estimation for when a task may terminate and therefore 
  	bound the execution time by the following two constraints 
	\begin{align}
		\varepsilon_{i, j} &\geq \sigma_{i,j} + \tBceT(i) + \iota_{i,j} 
			&& \forall i \in \mathcal{T}, 1 \leq j \leq m_i  \label{eq:epsilon0}\\
		\varepsilon_{i, j} &\leq \alpha_{i, j} + \tDead(i) 
			&& \forall i \in \mathcal{T}, 1 \leq j \leq m_i  \label{eq:epsilon1}
	\end{align}
	where the first constraint enforces that the task instance is considered at least for its best case 
	execution time, and the second constraints limits its execution time by the deadline of the
	according task. 
  	
  	However, if worst case execution times (WCET) or worst case response times (WCRT) are given, 
  	these would enable us to consider sporadic overload. These information 
    can easily be incorporated using the following constraints: 
  	\begin{align*}
	  	\varepsilon_{i, j} &\leq \alpha_{i, j} + WCRT(i)                 &&\forall j\colon 1 \leq j \leq m_i \\
	  	\varepsilon_{i, j} &\leq \sigma_{i, j} + WCET(i) + \iota_{i, j}  &&\forall j\colon 1 \leq j \leq m_i 
  	\end{align*}
  	Again, due to the focus on early development stages these constraints are
  	not considered in the tests of Section~\ref{sec:evaluation} but the 
  	implications of their usage are discussed in Section~\ref{sec:future}.

  	\subsection{Task Sequences}
  	A task sequence here is a finite, ordered sequence of tasks 
  	where a signal is interchanged from one task to another to implement 
  	a system function. It is important that these signals are pairwise 
  	causally related. 
 	On a more detailed level, an intra-ECU cause effect chain can be 
 	described in terms of an ordered sequences of so-called \emph{Executable Entities}
 	\cite{AUTOSAR-TimEx}. 
  	However these entities are used as an abstraction for executable code and 
  	are assignable to a task, which means that, in order to analyze cause-effect 
  	chains on this level of detail, it is sufficient to 
  	be able to analyze task sequences on instantiation-level. 
  	The timing properties of such a communicating task sequence again can be
  	characterized by an ordered sequence of points of time which describe when 
  	the value of a variable
  	is possibly processed at the corresponding task and when a response was calculated
  	(cf. \cite{Feiertag2008}). The response time of such a sequence then is the difference in time
  	between the point of time at which an instance of the first task could possibly
  	process the first signal and the point of time at which an instance of the last
  	task provides a response. 
  	For functionalities which are computed e.g. only every second task instance, 
  	we need to add a task a second time to the chain in order to obtain safe bounds. 
  	Since a maximum delay between two consecutive activations of the first 
  	task on the chain is given, we can give a bound on the time between an update of 
  	a variable and its next processing in a task instance. Therefore we are  
  	able to change the problem to finding the difference in time between the activation 
  	of the first task and the point of time at which the last task calculated the 
  	response. To obtain a safe bound when starting at an arbitrary point of time, we add
  	the maximum temporal gap between two consecutive task instances of the first 
  	task on the chain. 
  	
  	Corresponding to this definition, in our set of constraints
  	chains are sequences $p_{k=1}^\ell$ where $p_k\in\mathcal{T}$ f.a. 
  	$k\in\set{1,\dots,\ell}$ and $\ell\in\mathbb{N}_{\geq 1}$. Their 
  	encoding in the set of constraints is discussed below. 
  	The worst case end-to-end latency for such a chain is the maximum 
  	difference between the activation of the first task and the response
  	of the last task under the above-mentioned constraints. 	
  	This difference mainly depends on the relative offsets of 
  	the task pairs $p_{k}$ and $p_{k+1}$ f.a. $k\in\set{1,\dots,\ell-1}$.  
  	To get a safe upper bound for the response time 
  	we therefore need to ensure that all possible relative offsets 
  	are considered. We need to give an estimation for the lower 
  	bound of the length of the timespan in which what we call the 
  	\emph{critical offsets} will certainly occur. This 
  	timespan is discussed in Subsection~\ref{sec:relevantPeriod}. 	
  	For now let ${t^{\text{P}_{\text{min}}}}$ denote the length of this  
  	interval when starting at time zero. 
  	Given this interval the amount of instances for each task can be 
  	bounded with the minimum distance between two occurrences. 
  	Now, a bound for two values can be given: the indices which 
  	need to be considered for each task and the points of time 
  	at which accesses on the shared variables happen. Since 
  	all possible execution times are considered the latter 
  	is guaranteed to happen within $[0,{t^{\text{P}_{\text{min}}}}]$.
  	
  	For the problem of estimating the end-to-end latency of a task 
  	chain $p_{k=1}^\ell$ we add two additional sets of variables to our 
  	constraint program: $n$ and $x$. We use $n$ for the index 
  	of the task instance which participates in the chain at 
  	index $k$ 
  	and $x$ to hold the point of time at which the same instance 
  	has written its results certainly f.a. $k\in\set{1,\dots,\ell}$. 
  	The constraints 
  	on $n$ and $x$ 
  	$n_1 \geq 1$, $x_1\geq 0$ and are
  	f.a. $k\in\set{2,\dots,\ell}$ where $i$ is the index of $p_k$ and 
  	$i$ communicates recording the the implicit communication: 
  	\begin{align}
  		n_k &= \min\left(\set{j | \sigma_{i,j} \geq x_{i-1}}\right) \\
  		x_k &\geq \sigma_{i,n_k} \\
  		x_k &\leq \varepsilon_{i,n_k}\label{eq:implicitCommunication}
  	\end{align}
  	\autoref{eq:implicitCommunication} shows that in the case of 
  	implicit communications the consideration of more details regarding 
  	the actual execution times is promising, since here only the delay is 
  	relevant. However, in the case of explicit communication the BCET also 
  	gets relevant. Here, the constraints change f.a.  
    $k\in\set{2,\dots,\ell}$ where $i$ is the index of $p_k$ to: 
  	\begin{align}
  		n_k &= \min\left(\set{j | \sigma_{i,j} \geq x_{i-1}}\right) \\
		x_k &= \varepsilon_{i,n_k} \label{eq:explicitCommunication}
  	\end{align}  	
  	Lastly, in the case of deterministic communication the constraints f.a. 
  	$k\in\set{2,\dots,\ell}$ where $i$ is the index of $p_k$ change to: 
  	\begin{align}
  		n_k &= \min\left(\set{j | \alpha_{i,j} \geq x_{i-1}}\right) \\
		x_k &= \left(n_{k}+1\right)\cdot \tPeriod(p_k)\label{eq:detCommunication}
  	\end{align}

  	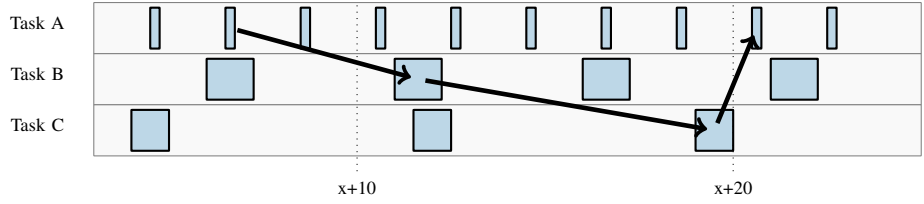
\begin{figure*}[tb]
  		\centering
  		\vspace*{0.25\baselineskip}\scriptsize
  		\begin{tikzpicture}[xscale=0.50,yscale=0.68]
  		
  		\node[] at (21.5,-0.5) (1er) {Task A}; 
  		\fill[rounded corners=.2pt, draw=gray, fill=gray!5!white]
  		(23.0,-1.1) rectangle ++(22.0, 1.0);
  		
  		\node[] at (21.5,-1.5) (1er) {Task B}; 
  		\fill[rounded corners=.2pt, draw=gray, fill=gray!5!white]
  		(23.0,-2.1) rectangle ++(22.0, 1.0);
  		
  		\node[] at (21.5,-2.5) (1er) {Task C}; 
  		\fill[rounded corners=.2pt, draw=gray, fill=gray!5!white]
  		(23.0,-3.1) rectangle ++(22.0, 1.0);

  		\foreach \x [count = \xi] in 
  		{  24.5, 26.5, 28.5, 30.5, 32.5, 34.5, 36.5, 38.5, 40.5, 42.5  }
  		{	             
  			\fill[thick, rounded corners=.2pt, draw=black, fill=fg0light]
  			(\x,-1) rectangle ++( 0.25, 0.80) node[pos=0.5] (2er\xi) { }; 
  		}	
  		
  		\foreach \x [count = \xi] in { 26.0, 31.0,  36.0, 41.0 }
  		{	             
  			\fill[thick, rounded corners=.2pt, draw=black, fill=fg0light]
  			(\x,-2) rectangle ++( 1.25, 0.80) node[pos=0.5] (5er\xi) { }; 
  		}	
  		
  		\foreach \x [count = \xi] in {  24.0, 31.5, 39.0 } 
  		{	             
  			\fill[thick, rounded corners=.2pt, draw=black, fill=fg0light]
  			(\x,-3) rectangle ++( 1.00, 0.80) node[pos=0.5] (7er\xi) { }; 
  		}

  		\foreach \x in {30,40} {
  			\draw[dotted] (\x,-3.4) -- (\x,-0.1) node[pos=-0.1] (\x)
  			{x+\pgfmathparse{int((\x-20)}\pgfmathresult}; 
  		}
  		
  		\tikzstyle{chain} = [line width=1.75pt, color=black, ->]
  		  		
  		\draw[chain, solid]          (2er2) -- (5er2);
  		\draw[chain, solid]          (5er2) -- (7er3);
  		\draw[chain, solid]          (7er3) -- (2er9);
  		
  		\end{tikzpicture}
  		\caption{Estimation neglecting actual execution}
  		\label{fig:greedyEstimation}
  	\end{figure*}
  
  	\begin{figure*}[tb]
		\centering
		\vspace*{0.25\baselineskip}\scriptsize
		\begin{tikzpicture}[xscale=0.50,yscale=0.68]
		
		\node[] at (21.5,-0.5) (1er) {Task A}; 
		\fill[rounded corners=.2pt, draw=gray, fill=gray!5!white]
		(23.0,-1.1) rectangle ++(22.0, 1.0);
		
		\node[] at (21.5,-1.5) (1er) {Task B}; 
		\fill[rounded corners=.2pt, draw=gray, fill=gray!5!white]
		(23.0,-2.1) rectangle ++(22.0, 1.0);
		
		\node[] at (21.5,-2.5) (1er) {Task C}; 
		\fill[rounded corners=.2pt, draw=gray, fill=gray!5!white]
		(23.0,-3.1) rectangle ++(22.0, 1.0);

		\tikzstyle{exec} = [thick, rounded corners=.2pt, draw=none, fill=fg0];
		
		\foreach \x [count = \xi] in 
		{  24.5, 26.5, 28.5, 30.5, 32.5, 34.5, 36.5, 38.5, 40.5, 42.5  }
		{	             
			\fill[thick, rounded corners=.2pt, draw=black, fill=fg0light]
			(\x,-1) rectangle ++( 0.25, 0.80) node[pos=0.5] (2er\xi) { }; 
		}	
	
		\foreach \x [count = \xi] in 
		{  24.50, 26.50, 28.50, 30.50, 32.50, 34.50, 36.50, 38.50, 40.50, 42.50  }
		{	             
			\fill[exec] (\x,-0.25) rectangle (\x+0.2,-0.95);
		}
		
		\foreach \x [count = \xi] in { 26.0, 31.0,  36.0, 41.0 }
		{	             
			\fill[thick, rounded corners=.2pt, draw=black, fill=fg0light]
			(\x,-2) rectangle ++( 1.25, 0.80) node[pos=0.5] (5er\xi) { }; 
		}	
	
		\fill[exec] (26.025,-1.25) rectangle (26.500,-1.95);
		\fill[exec] (31.025,-1.25) rectangle (31.500,-1.95) node[pos=0.5] (5er) { };
		\fill[exec] (36.025,-1.25) rectangle (36.500,-1.95);
		\fill[exec] (41.025,-1.25) rectangle (41.400,-1.95);

		\foreach \x [count = \xi] in {  24.0, 31.5, 39.0 } 
		{	             
			\fill[thick, rounded corners=.2pt, draw=black, fill=fg0light]
			(\x,-3) rectangle ++( 1.00, 0.80) node[pos=0.5] (7er\xi) { }; 
		}
		
		\fill[exec] (24.025,-2.25) rectangle (24.500,-2.95);
		\fill[exec] (31.500,-2.25) rectangle (32.050,-2.95) node[pos=0.5] (7er) { };
		\fill[exec] (39.025,-2.25) rectangle (39.550,-2.95);
		
		\foreach \x in {30,40} {
			\draw[dotted] (\x,-3.4) -- (\x,-0.1) node[pos=-0.1] (\x)
			{x+\pgfmathparse{int((\x-20)}\pgfmathresult}; 
		}
		
		\tikzstyle{chain} = [line width=1.75pt, color=black, ->]
		
		\draw[chain, solid] (2er2) -- (5er);
		\draw[chain, solid] (5er) -- (7er.west);
		\draw[chain, solid] (7er.east) -- (2er5);
		
		\end{tikzpicture}
		\caption{Estimation considering actual execution}
		\label{fig:detailEstimation}
	\end{figure*}
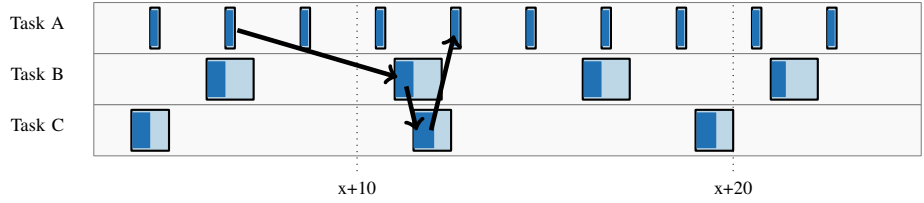  

	Before we discuss the interval which needs to be considered for safe 
	estimations, we want to give a motivational example for the 
	consideration of the actual processing times of each task. 
	Therefore,~\autoref{fig:greedyEstimation} and~\autoref{fig:detailEstimation} 
	give an example on the impact of neglecting the actual possibly 
	execution times of the task instances. The priorities are: 
	$\tPrio(\text{Task A}) > \tPrio(\text{Task B}) > \tPrio(\text{Task C})$. 
	We assume implicit communication and the following sequence:  
	$(\text{Task A}, \text{Task B}, \text{Task C}, \text{Task A})$. 
	In~\autoref{fig:greedyEstimation} the actual execution times are 
	unknown to the solver. We see that the it can not safely say that the 
	second instance of Task C has not started before the second instance 
	of Task B has written its results. \autoref{fig:detailEstimation} again
	shows the relative offsets of the tasks in~\autoref{fig:greedyEstimation}
	but with respect to actual execution times which are shown darker. 
	Since the priority of Task B is higher than the priority of Task C 
	it will always finish before the scheduler selects Task C for processing. 
	Therefore the solver can safely reason that the second instance of 
	Task C will always process the data of the second instance of Task B. 
	This makes the worst-case path depicted in \autoref{fig:greedyEstimation}
	infeasible. 

	\subsection{Relevant Period} \label{sec:relevantPeriod}
	\begin{figure}[t]
		\centering
		\vspace*{\baselineskip}
		\begin{tikzpicture}[xscale=1.00,yscale=1.00]
		
		\draw[->] ( 0.5, 0.0) -- ( 8.0, 0.0) node[right] {$t$};
		
		\draw[         ] ( 1.00, 0.10) -- ( 1.00,-0.10); 
		\draw[(-)      ] ( 1.00,-0.40) -- ( 2.50,-0.40); 
		\draw[draw=none] ( 1.00,-0.40) -- ( 2.50,-0.40) node[fill=white,midway,below=-.25] {offset};
		\draw[         ] ( 2.50, 0.10) -- ( 2.50,-0.10); 
		
		\draw[         ] ( 2.50, 0.10) -- ( 2.50,-0.10); 
		\draw[(-)      ] ( 2.50,-0.40) -- ( 5.00,-0.40); 
		\draw[draw=none] ( 2.50,-0.40) -- ( 5.00,-0.40) node[fill=white,midway,below=-.25] {$LCM$};
		\draw[         ] ( 5.00, 0.10) -- ( 5.00,-0.10); 
		
		\draw[         ] ( 5.00, 0.10) -- ( 5.00,-0.10); 
		\draw[(-)      ] ( 5.00,-0.40) -- ( 7.50,-0.40); 
		\draw[draw=none] ( 5.00,-0.40) -- ( 7.50,-0.40) node[fill=white,midway,below=-.25] {$UB$};
		\draw[         ] ( 7.50, 0.10) -- ( 7.50,-0.10); 
		
		
		\draw[dashed   ] ( 4.50, 0.00) -- ( 4.50, 0.45); 
		\draw[(-)      ] ( 4.50, 0.45) -- ( 6.90, 0.45); 
		\draw[draw=none] ( 4.50, 0.45) -- ( 6.90, 0.45) node[fill=white,midway,below=-.25] {\scriptsize 
			possible sequence};
		\draw[dashed   ] ( 6.90, 0.00) -- ( 6.90, 0.45); 
		
		\draw[dashed   ] ( 1.00, 0.00) -- ( 1.00, 0.95); 
		\draw[(-)      ] ( 1.00, 0.95) -- ( 7.50, 0.95); 
		\draw[draw=none] ( 1.00, 0.95) -- ( 7.50, 0.95) node[fill=white,midway,below=-.25] {relevant period};
		\draw[dashed   ] ( 7.50, 0.00) -- ( 7.50, 0.95); 
		\end{tikzpicture} 
		\caption{Relevant interval in context}
		\label{fig:relevantPeriod}
	\end{figure}
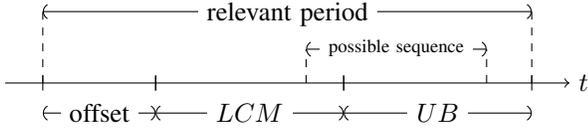  	

	The constraints stated in the previous sections describe a schedule in which 
	each task $i$ is considered with $m_i$ task instances, i.e. a schedule 
	in a certain interval of time is described. 
	It is crucial that this interval is chosen sufficiently large to ensure a 
	sound computation of upper bounds on the response times of task chains. 
	On the other hand, the solving time increases with larger intervals, thus it is preferable 
	to consider intervals which are not larger then necessary. 
	\autoref{fig:relevantPeriod} sketches the interval we 
	used for our analyses. 
	We first describe its derivation, and afterwards discuss that not all tasks must actually 
	be considered, i.e. the tasks which are relevant for the size of the interval. 
	
	Let $LCM$ denote the least common multiple of periods of tasks which are relevant for the size of the interval. 
	A worst-case occurrence of a task chain will start at some point of this so-called hyper period, 
	thus the interval for the analysis must contain a full hyper period. Next, the interval must be large enough such that
	this occurrence fits into it, independent of its starting time. 
	We thus extend the interval by UB, a trivial upper bound on the length of the chain, as shown on the right hand side of Figure~\ref{fig:relevantPeriod}. 
	Such bounds can be derived by summing up trivial upper bounds on the worst case response time of every task on the chain. 
	As the execution of task instance may be delayed due to the execution of other tasks, we add an offset 
	to the interval size (on the left-hand side of Figure~\ref{fig:relevantPeriod}) such that every task can be started at least once before this interval. 
	For periodic tasks, this requires an offset of $O_p = \max\set{\tOffset(i) + \tPeriod(i)| i \in \mathcal{T}}$. 
	In the case of sporadic tasks, let $O_s$ represent the maximum offset between two occurrences of these tasks. 
	Then we choose the \emph{relevant interval} 
	as the interval $[0, T]$ with $T=O+LCM+UB$ and $O=\max\{O_s,O_p\}$.
	
	Assume this interval was not sufficiently large, i.e. there was a $T' > T$ such that choosing the interval $[0, T']$ would yield a larger end-to-end delay, and that $T'$ is minimal with this property. 
	Then, the chain must start at a time larger than $O + LCM$. Subtracting $LCM$ from all values for the variables $\alpha, \sigma, \varepsilon$ and adjusting the right-hand side of the constraints accordingly yields a shifted solution. 
	Now, all task instance occurring at a negative point of time can be ignored. If they had a influence on the solution,  
	using the offset on the left-hand side allows for simulating the according system load. This yields a new solution within a smaller interval, contradicting the assumption that $T'$ was minimal. 
	
	Real-world problem instances often contain tasks which run with a large period, e.g. $1$ second, and a very low priority. As such tasks significantly increase the hyper period, we seek to focus on relevant tasks, i.e. tasks which actually have an influence on the end-to-end delay of a task chain. 
	
	We therefore define the set of relevant tasks %
		$\mathcal{T}_{\text{rel}}$ for 
		a chain $p_{k=1}^\ell$ as the smallest subset of $\mathcal{T}$
		such that: 
		\begin{enumerate}
			\item All tasks which occur on the chain are relevant: For all $t\in\mathcal{T}$, 
			if exists $k\in\set{1,\dots,\ell}$ with 
			$p_k=t$, then $t\in\mathcal{T}_{\text{rel}}$.  
			\item Tasks which have a higher priority than a relevant task which runs on the same 
			core also become relevant: For all $t\in\mathcal{T}$, 
			if exists $t'\in\mathcal{T}_{\text{rel}}$ such that 
			$\tCore(t') = \tCore(t)$ and $\tPrio(t) > \tPrio(t')$, 
			then $t\in\mathcal{T}_{\text{rel}}$.  
			\item Non-preemptable task may have an influence on the task chain. 
			For all $t\in\mathcal{T}$, 
			if exists $t'\in\mathcal{T}_{\text{rel}}$ such that 
			$\tCore(t') = \tCore(t)$ and $\tPree(t) = \mathtt{f}$,
			then $t\in\mathcal{T}_{\text{rel}}$.  
			\item Tasks which activate relevant tasks must also be considered. 
			For all $t\in\mathcal{T}$ where $\tActT(t)=\texttt{chained}$, 
			if exists $t'\in\mathcal{T}_{\text{rel}}$ such that 
			$t'$ is chained to $t$, then $t\in\mathcal{T}_{\text{rel}}$.  
			
		\end{enumerate}
		This is, $\mathcal{T}_{\text{rel}}$ contains the tasks which occur on the chain, 
		and every task which may influence a relevant task. 
		
		As we will show in the next section, considering only relevant tasks for the computation 
		of the interval size can reduce its size by an order of magnitude. 

  	\section{Evaluation} \label{sec:evaluation}
 	We encoded the constraints described in the previous section 
 	in \emph{MiniZinc}~\cite{DBLP:conf/cp/NethercoteSBBDT07}.
 	This allows for both a structured representation of our problem and for using 
 	a wide range of solvers as backend. 	
	\emph{MiniZinc} models together with data describing a particular problem instance are 
	translated to a \emph{FlatZinc} formula, which is then solved by a solver backend.
	In our experiments we chose \textsc{Chuffed} with parallelization 
	as presented in \cite{DBLP:conf/cpaior/EhlersS16} as the solver. 

 	The solver initially has no indications on how to branch on the different variables. 
 	Therefore, in order to decrease the time needed to find a result, we use so-called
 	search annotations to impose a search strategy by influencing branching decisions.
 	Generally, we add annotations telling the solver to determine the activation of a task before 
 	trying to determine its termination by branching on a small value for $\sigma$ and 
 	a large value for $\varepsilon$. Furthermore we ensure that periodic tasks are 
 	put in order first and chained tasks last. 
 	In this way, the solver is led to parts of the search space which describe situations 
 	with high system load first, as it is more likely to find a large response time here.

	\begin{table*}[t] 
	\caption{Performance of the approach}
	\label{tbl:Performance}
	\begin{center}
		\begin{tabular}{l l r r r r r r}
			\toprule
			\multicolumn{1}{p{2.80cm}}{{\textbf{Method}}} & 
			\multicolumn{1}{p{1.80cm}}{{\textbf{Task set}}} & 
			\multicolumn{2}{p{2.10cm}}{\centering{\textbf{Compiler}}} & 
			\multicolumn{1}{p{0.20cm}}{\centering{                 }} & 
			\multicolumn{2}{p{2.10cm}}{\centering{\textbf{Solver}}}   & 
			\multicolumn{1}{p{1.50cm}}{\centering{\textbf{Latency}}}  \\
			&                             &       s &   kbyte &&      s  &  kbyte  &  $\mu$s \\ 
			\cmidrule(l{.75em}){3-4} \cmidrule(l{1.75em}){5-7}
			Full period           & PTC~A &    0.93 &   38872 &&    0.29 &   44280 &   47000 \\  
			Full period           & PTC~B &    2.69 &  167644 &&    1.44 &  104744 &   17250 \\
			Full period           & ECM   &    1.41 &   87824 &&    1.55 &  180088 &   19146 \\[0.6ex] 
			Decomposition         & PTC~A &       - &       - &&       - &      -  &       - \\  
			Decomposition         & PTC~B &    1.54 &   98940 &&    0.64 &   71284 &   17250 \\
			Decomposition         & ECM   &    0.92 &   56704 &&    0.69 &   99700 &   19146 \\[0.6ex] 
			\bottomrule
		\end{tabular}
		\normalsize
	\end{center}
	\end{table*}

	Furthermore, we use a preprocessing step to simplify many constraints. 
	For example, in the sum in~\autoref{eq:iota} it is sufficient to consider
    only task instances which might possibly interact with the task instance $(i, j)$. 
    In many cases, this significantly reduces the number of summands and the time 
    required for translating the MiniZinc problem into FlatZinc.
	Additionally, with an eye to increasing complexity due to an
    increased level of detail, we introduce two variations of the problem.
	Firstly, computing time can be reduced by splitting the relevant period 
	into smaller, overlapping intervals and estimate response times for each 
	of these intervals separately. The results of this analysis are not 
	safe but they give a lower bound. 
	Secondly, an upper bound for the maximum response time of a task chain can be 
	computed by relaxing the problem. This might be interesting when adding more 
	complexity to the task set or when a very fast evaluation of different configurations
	needs to be performed. Ignoring the constraints~\ref{eq:sigma} 
	and~\ref{eq:iota} massively simplifies the problem, and yields interesting
	bounds in our experiments. However, for the considered level of detail 
	the full period with all constraints can be solved on a general-purpose 
	computer. \todo{Relax noch erstmal noch drin!}
		
	This has been tested with three different industrial-scale task sets. 
	Two of them model control units which can be found in the Daimler powertrain (PTC~A and PTC~B). 
	The	third one was taken from the \emph{FMTV 2016 Verification Challenge} of Bosch (ECM) with 
	small adjustments. It is a realistic example of an engine control unit. The details of task sets are: 
	\begin{description}
		\item[PTC~A]~
		 Is taken from an ECU with two cores which runs 17 task containers. 
		 Nine tasks are activated periodically on the first core, 
		 eight run the second core and are chained to a counterpart 
		 on core one. The chain is an example of a chain from 
		 a networking task via middleware to application and back. 
		\item[PTC~B]~
		 Is taken from an ECU with four cores which runs 39 task containers. 
		 23 tasks are activated periodically with activation periods 
		 between 1ms and 1000ms. The other tasks are executed 
		 sporadically. We made assumptions for the minimum time 
		 distance of two consecutive occurrences. 
		 Again the chain is an example of a chain from 
		 a networking task via middleware to application and back. 
		\item[ECM]~
	     The task set of the ECU described in \cite{Hamann16}. 
	     Unfortunately the task-level chains in the provided 
	     data are rather short. Thus, we designed a more complex chain here. 
	     We assumed functionalities for the tasks and then created a chain which
	     spans over six tasks. It is intended to represent a path from angle synchronous 
	     software to an application software and back.  
	\end{description} 
	
	The calculations were performed on a notebook equipped with an 
	Intel Core i7-3740QM CPU, 16GB of RAM, running Ubuntu 16.04 LTS. We 
	used the parallel version of chuffed on $4$ cores. The results are shown 
	in~\autoref{tbl:Performance}. For the "Full period"  we considered an 
	interval of 
	$111 ms$ for the PTC~A,
	$41 ms$ for the PTC~B, and 
	$40 ms$ for the ECM. 
	For the decomposition of the PTC~A we then used the 
	same interval also for the decomposition, resulting in only 
	one interval to check. For the 
	PTC~B and ECM we split the \emph{relevant period} in 
	$2$ equal-sized time slices of $30000\mu s$. 
	The computational resources needed until the result from "Full period" was 
	found are shown in the rows "Decomposition". 
	For the benchmarks PTC~B and ECM with the analyzed chains, this always
	happened within the first time slice. However, this is 
	in general not the case, e.g. when size or overlapping of the 
	intervals are chosen disadvantageously. 
	Lastly, in the case of differences in the
	results of the relaxed and the detailed model, 
	the precision 
	of latency estimation without the knowledge of 
	execution times has been improved. In our experiments 
	we saw differences of up to $5 ms$ which means a clear 
	improvement of the quality of the estimation as it is a
	whole period of a task in the analyzed task sequence. 
	\todo{Can we specify this? <- not with much details}
	\section{Future Work} \label{sec:future}%
  	In this paper, we presented an approach focused on early development 
  	stages. Future work takes aim at later development stages, where it  
  	is possible to integrate further constraints based 
  	on data which usually becomes available in later development stages.
  	Worst case execution times or worst case response times may be 
  	available then. 
  	This would on the one hand enable us to consider overload 
  	situations, on the other hand intervals in which data propagation is possible
  	could be bounded more precisely.
  	Furthermore, one may obtain tighter bounds on the response time 
  	of the last task in the chain. 
  	These possible advantages are in contrast with new challenges. 
  	Some overload situations may be infeasible, e.g. in cases where the WCETs of certain tasks cannot  
  	happen simultaneously. 
  	Additionally, overload situations need to be bound somehow, otherwise the 
  	solver will use them to stretch response times to infinity.   	
  	Hence, going one level deeper to do module-level analyses seems promising 
  	although not trivial. 
  	
 	\section{Conclusions} 
 	In this paper we presented a template for a constraint program
 	to estimate end-to-end latencies of task sequences in task sets 
 	of multi-core ECUs. 
 	This can be used to obtain a safe estimation for 
 	the response time 
 	of task sequences within the task set. 
 	For large scale systems, relaxation and decomposition can be used to compute 
 	lower and upper bounds on the response time. 
 	For the basic approach very few information about the system to analyze 
 	is needed, nevertheless our approach would allow to add more 
 	details about the task containers, if available. 
 	The results can be used to express the abstract timing 
 	behavior of an ECU for arriving network messages. This is done by analyzing 
 	the part of the system-level cause effect chain which starts at the 
 	receiving network task of the ECU and ends at the corresponding network task 
 	to propagate the stimulus or a task controlling an actuator. 
 	We showed how to utilize a parallel solver for this 
 	specific problem and the practical application shows 
 	that the approach scales for industrial-size problems while running 
 	on a general-purpose computer. 

	\bibliographystyle{IEEEtran}
	\bibliography{IEEEabrv,99_lit}

\end{document}